\documentclass[twocolumn,aps,amsmath,amssymb,floatfix,prl,showkeys,superscriptaddress]{revtex4}

\usepackage{bbm}
\usepackage[dvips]{graphicx}
\usepackage{amsmath,amssymb,latexsym}
\usepackage{enumerate}
\usepackage{xcolor} 

\newcommand{\Drot}{D_\mathrm{rot}}
\newcommand{\Deff}{D_\mathrm{eff}}

\newcommand{\R}{\mathbf{R}}

\newcommand{\x}{\mathbf{x}}

\newcommand{\tauopt}{\tau_\mathrm{opt}}

\renewcommand{\ol}[1]{{\overline{#1}}}

\begin{document}

\title{Composite search of active particles in three-dimensional space\\ based on non-directional cues} 

\author{Justus A. Kromer}
\affiliation{Stanford University, Stanford, California, United States of America}
\author{Andrea Auconi}
\affiliation{TU Dresden, Dresden, Germany}
\author{Benjamin M. Friedrich}
\email{benjamin.m.friedrich@tu-dresden.de}
\affiliation{TU Dresden, Dresden, Germany}

\date{\today}


\begin{abstract} 
We theoretically address minimal search strategies of active, self-propelled particles towards hidden targets in three-dimensional space.
The particles can sense if a target is close, 
e.g., by detecting signaling molecules released by a target, 
but they cannot deduce any directional cues. 
We focus on composite search strategies, where particles switch between extensive outer search and intensive inner search;
inner search is started when the proximity of a target is detected and ends again when a certain inner search time has elapsed.
In the simplest strategy, active particles move ballistically during outer search, 
and transiently reduce their directional persistence during inner search.
In a second, adaptive strategy, particles exploit a dynamic scattering effect by reducing directional persistence only outside a well-defined target zone.
These two search strategies require only minimal information processing capabilities and a single binary or tertiary internal state, respectively,
yet increases the rate of target encounter substantially.
The optimal inner search time scales as a power-law with exponent $-2/3$ with target density, 
reflecting a trade-off between exploration and exploitation.
\end{abstract}

\maketitle
Turning active colloids into micro-scale robots 
will enable envisioned biomedical or environmental applications \cite{Li2017,Jurado2018}.
Yet, the information processing capabilities of these self-propelled agents remain limited.
This motivates theoretical analysis of minimal strategies that are simulataneously simple and effective.
Here, we address optimal random search for hidden targets \cite{Mijalkov2016,Nava2017}. 
We propose a minimal model of an Active Brownian Particle (ABP) that regulates its directional persistence in response to local cues.
Previous work showed that for ABP without internal states or memory, such a chemokinesis strategy does not provide any advantage
compared to simple ballistic motion \cite{Kromer2020}. 
Yet, a single binary or tertiary internal state allows these agents to increase their rate of target encounter substantially.

Our question is thus:
\textit{How to continue a random search for hidden targets
if information is received that a target must be close,
yet no directional information is given?}
This situation frequently arises when search agents detect dilute chemical signals released by targets.
Examples include chemokinetic navigation of motile cells, 
foraging of animals \cite{Fraenkel1940},
and odor-sniffing robots \cite{Webb2002,Settles2005}. 
For micrometer-sized agents, 
the signal-to-noise ratio of sensing chemical gradients is low at sub-nanomolar concentrations, 
preventing effective directed chemotaxis, 
nonwithstanding the fact that the chemical signal itself is above detection threshold, 
and thus indicates the proximity of a target releasing signaling molecules.

If search agents can only detect the presence of a nearby target within a given threshold distance, 
we can decompose the search problem into an \textit{outer} and an \textit{inner} search problem, 
far and close to a target, respectively.
While the main purpose of outer search is the \textit{exploration} of search space (extensive search \cite{Benhamou2014}),
the agent may choose a different \textit{exploitation} strategy in the vicinity of a target for inner search (intensive search).
Composite search strategies allow agents to switch between different search modes 
\cite{Benhamou1992,Plank2008,Benichou2011,Benhamou2014,Bartumeus2014,Nolting2015,Palyulin2016},
e.g., for outer and inner search.

The performance of search strategies strongly depends on the dimension of search space. 
According to P{\'o}lya's theorem \cite{Polya1921}, 
a diffusive particle will eventually find a target in one- and two-dimensional space almost surely,
while such a particle may never find the target in three dimensional space. 
A colloquial version of P\'olya's theorem states:
\textit{A drunk man will find his way home, but a drunk bird may get lost forever} \cite{Kakutani}.
This marks an important qualitative difference between search problems in two and three space dimensions.
Previous research often investigated search problems in two-dimensional space, largely motivated by animal foraging. 
Among others, this includes theoretical and computational studies of composite search strategies for hidden targets 
\cite{knoppien1985predators, benhamou2007many, reynolds2009adaptive, Nolting2015, Schwarz2016}, 
the narrow escape problem \cite{Schuss2007, benichou2008narrow, Schwarz2016, mangeat2019narrow}, 
or local search around a home base \cite{noetel2018optimal, noetel2018search}.
Yet, microswimmers must usually search a three-dimensional environment. 
Therefore, in the present manuscript, we focus on the less-studied problem of search in three-dimensional space.

The optimal strategy for outer search depends on the distribution of target sites:
while L\'evy walks are optimal for certain fractal distributions of targets (or if targets can be revisited), 
simple ballistic motion is optimal in infinite domains with uncorrelated random target positions, 
where each target can be visited only once
\cite{Viswanathan1999, Bartumeus2002, Tejedor2012}.
This latter case exactly corresponds to a well-stirred suspension of targets
(which are consumed upon encounter), 
and we will therefore assume ballistic motion for outer search.

Directional persistence is a key kinetic parameter of self-propelled microswimmers, 
which is set by a competition between swimming speed and 
directional fluctuations; 
directional fluctuations can be of thermal origin or originate from active fluctuations of the propulsion mechanism itself.
A microswimmer may thus \textit{control} its directional persistence 
by different means:
(i) transiently changing its speed \cite{Vuijk2018}, 
for instance in response to the local concentration of a chemical \cite{peng2015self} or external fields \cite{zhang2009characterizing},
(ii) transiently changing its effective size $L$ 
(since the thermal rotational diffusion coefficient changes as $\sim1/L^3$ \cite{Einstein1906}), or
(iii) up- or down-regulating any stochastic part of active propulsion \cite{Berg1972,Ma2014}. 
Despite these different possible implementations, 
we will consider a generic model, assuming agents can effectively regulate their directional persistence. 

In this short communication,
we consider a minimal model of an active particle
that employs composite search in three-dimensional space. 
The particle switches between ballistic motion in outer search and intensive inner search.
Our model generalizes Active Brownian Particles (ABP),
frequently used as minimal model for cell motility, 
e.g.\ of biological or artificial microswimmers \cite{Romanczuk2012, BechingerRevModPhys}
(also called correlated or persistent random walkers \cite{Gillis1955,Masoliver1989,Friedrich2008}). 
We find that a single binary internal state allows these agents to substantially increase their rate of target encounter.
The optimal inner search time (`giving-up time') decreases with target density $\rho$ as $\sim\rho^{-2/3}$, 
reflecting a exploration-exploitation trade-off \cite{Reddy2016} between
continuing inner search to eventually find the nearest target, or searching elsewhere for other targets.

\begin{figure}
\includegraphics[width=\linewidth]{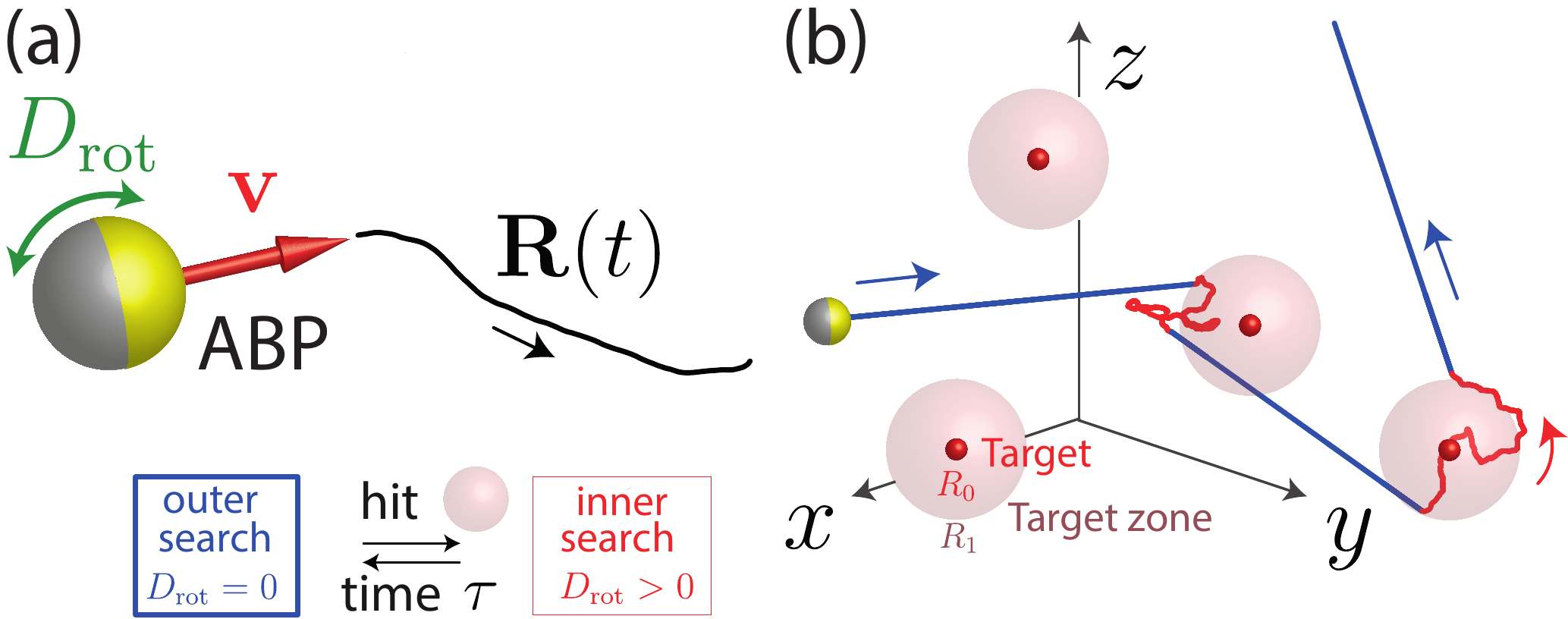}
\caption{
\textbf{Composite search.} 
(a) We consider an Active Brownian Particle (ABP) with velocity $v$ and rotational diffusion coefficient $\Drot$ 
as a generic model of a self-propelled microswimmer.
(b) The ABP searches for randomly distributed hidden targets (red) in three-dimensional space, 
alternating between an \textit{outer search} (blue, here corresponding to ballistic motion), 
and \textit{inner search} for a restricted time $\tau$, (red, here corresponding to low-persistence motion).
Inner search is triggered whenever the ABP senses the vicinity of a target (ros\'e), 
and ends once a certain inner search time $\tau$ has elapsed. This defines a composite search strategy based on non-directional cues.
}
\label{figure1}
\end{figure}

\paragraph{Model: Active Brownian Particles.}

We consider an ABP moving along a trajectory $\textbf{R}(t)$ in three-dimensional space
with speed $v$ and rotational diffusion coefficient $D_\mathrm{rot}$.
Rotational diffusion causes its tangent $\mathbf{t}=\dot{\mathbf{R}}/v$ 
to decorrelate on a time-scale $\tau_p=l_p/v$ set by the persistence length $l_p=v/(2 D_\mathrm{rot})$, 
i.e.,
$\langle \mathbf{t}(t_0) \cdot \mathbf{t}(t_0+t) \rangle = \exp ( -|t|/\tau_p)$
\cite{Daniels1952}.

We first consider the case of a single spherical target of radius $R_0$ located at $\R{=}\mathbf{0}$. 
Due to spherical symmetry,
the time-dependent distance $R(t)=|\mathbf{R}(t)|$ of the agent to the origin, 
and the time-dependent angle $\psi(t)$ enclosed by the tangent $\mathbf{t}$ 
and the radial direction $\mathbf{e}_R=-\mathbf{R}/R$, 
decouple from other coordinates \cite{Friedrich2007,Friedrich2009,Kromer2020}
\begin{eqnarray}
\label{eq:rPsiDynamics1}
\dot{R} &=& - v \cos\psi, \\
\label{eq:rPsiDynamics2}
\dot{\psi} &=& \frac{v}{R}\, \sin\psi +\sqrt{2 \Drot}\, \xi(t) + \Drot\, \mathrm{cot}\,\psi.
\end{eqnarray}   
Here, $\xi(t)$ is Gaussian white noise with 
$\langle \xi(t) \rangle=0$ and $\langle \xi(t)\xi(t') \rangle=\delta(t-t')$. 

Intriguingly, 
the case of a position-dependent speed $v(\x)$ (\textit{orthokinesis} \cite{Fraenkel1940}), and 
the case of a position-dependent rotational diffusion coefficient $\Drot(\x)$ (\textit{klinokinesis}) 
can be mapped onto each other using a local re-parameterization of time \cite{Kromer2020}:
if we formally re-scale local time by a factor $\Phi=v_0/v(\x)$, 
then ABPs apparently move with constant speed $v_0=v(\x)\Phi$, 
while the new rotational diffusion coefficient reads $\Drot\Phi$. We emphasize that the shapes of trajectories do not change under such time re-parametrization. For ABP with constant speed, 
it is known that the steady-state probability distribution $p^\ast(\x)$ is spatially homogeneous, 
even if the rotational diffusion coefficient $\Drot(\x)$ depends on position \cite{Blanco2003,Benichou2005,Kromer2020}.  

In the following, we assume a minimal model with constant speed $v$ 
and switchable rotational diffusion coefficient $\Drot$, 
as detailed next. 

\paragraph{Inner and outer search.}

We now turn to the full search problem for multiple, non-revisitable targets
(Poisson distributed with density $\rho$), 
combining inner and outer search, see Fig.~\ref{figure1}.
In outer search mode,
the ABP moves ballistically ($\Drot=0$),
i.e., it uses the optimal strategy to find non-revisitable targets \cite{Viswanathan1999, Bartumeus2002, Tejedor2012}. 
Close to a target, the ABP uses time-restricted inner search with high rotational diffusion coefficient $\Drot$
(low directional persistence), 
see Fig.~\ref{figure1}.
This simple strategy only requires the agent to detect that a target must be within a detection radius $R_1$.

\begin{figure}
\includegraphics[width=\linewidth]{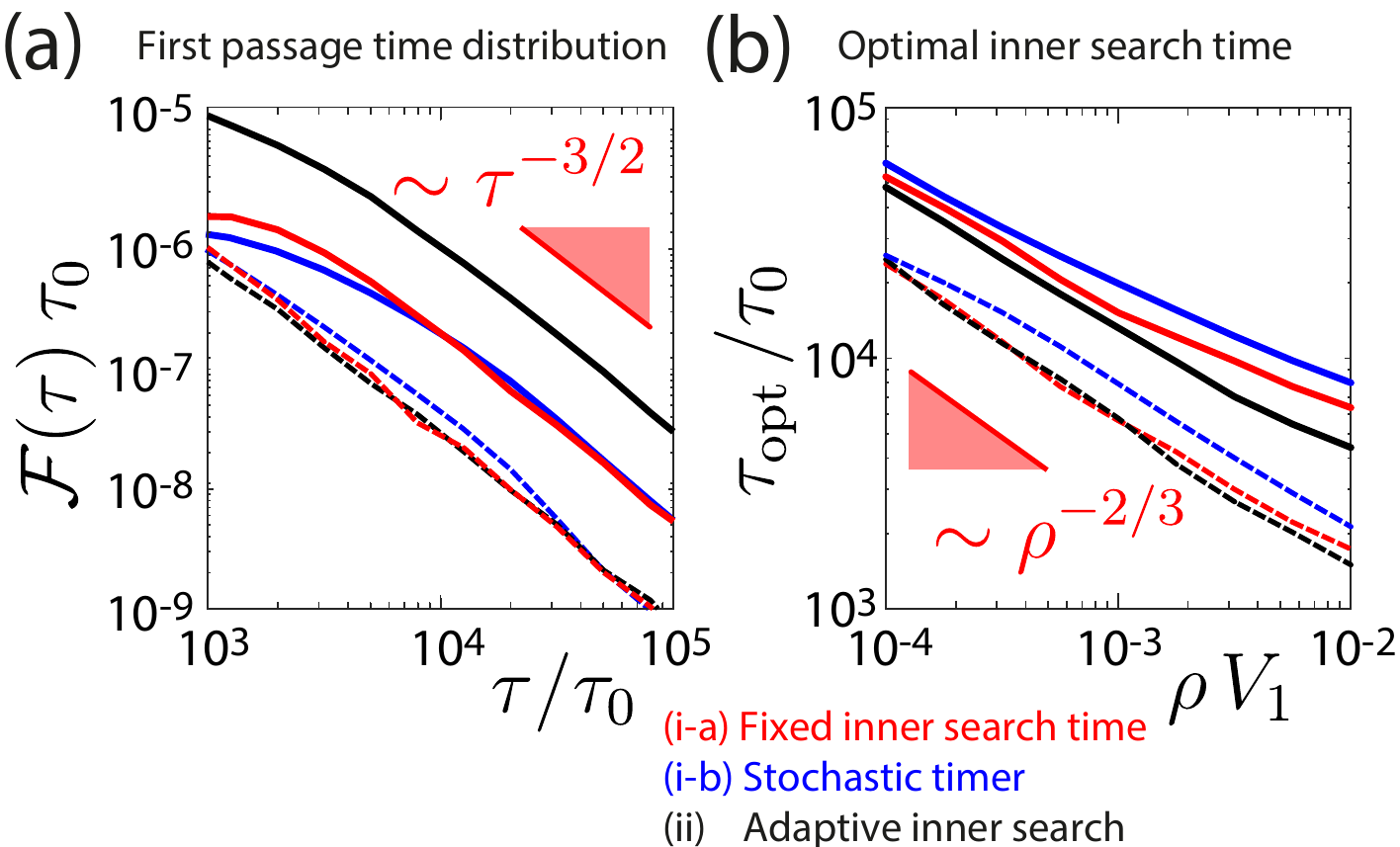}
\caption{
\textbf{Optimal inner search time decreases with target density.} 
(a): 
Simulated probability density 
$\mathcal{F}(\tau)=d p_\mathrm{in}(\tau)/d\tau$ 
of first-passage times $\tau$ to find a spherical target of radius $R_0$
for search agents starting at a distance $R_1$ from the target
for different variants of inner search.
Red curves: agents with constant rotational diffusion coefficient $\Drot$ in inner search [strategy (i-a)].
Black: agents employing adaptive inner search [strategy (ii)]
with ballistic motion inside the target zone $R\le R_1$ and constant rotational diffusion coefficient $\Drot$ outside, as in \cite{Kromer2020}.
Blue curves: probability density $\ol{\mathcal{F}}(\ol{\tau}) = d \ol{p}_\mathrm{in}(\ol{\tau})/d\ol{\tau}$
to find the target for agents with stochastic timer, 
for which maximal inner search time follows an exponential distribution with mean $\ol{\tau}$, 
see Eq.~(\ref{eq:qbar}) [strategy (i-b)].
All curves: dashed: $\Drot{=}v/(4R_0)$, solid: $\Drot{=}10\,v/(4R_0)$, time-scale $\tau_0=R_0/v$.
The probability densities are not normalized, 
as the total probability to eventually find the target is less than one even for infinite search times.
(b):
Optimal time limit $\tau_\mathrm{opt}$ for inner search decreases for increasing target density
[cases as in panel (a)].
Parameters: $R_1/R_0=20$, 
volume of target zone $V_1{=}4\pi R_1^3/3$.
}
\label{figure2}
\end{figure}

Whenever the ABP enters one of the spheres of radius $R_1$ surrounding the targets, it switches to one of the following \textit{inner search} strategies for a limited inner search time $\tau$:  
\begin{enumerate}[(i)]
\item[(i-a)] 
\textit{Inner search with fixed inner search time:} The ABP transiently increases its rotational diffusion coefficient $\Drot$ (i.e., reduces its directional persistence)
upon detecting a target zone
for a total duration $\tau$ (red curves in Fig.~\ref{figure2}). 
\item[(i-b)]
\textit{Stochastic timer}:
Same as strategy (i-a), but for random inner search times $\tau$
drawn from a distribution $q(\tau)$ with mean $\ol{\tau}$ (blue curves in Fig.~\ref{figure2}).
\item[(ii)]
\textit{Adaptive inner search}: 
During inner search, the ABP uses a position-dependent rotational diffusion coefficient $\Drot(\x)$
that depends on its distance $R$ from the nearest target: 
the agent uses ballistic motion with $\Drot=D_1=0$ if the target is within the detection radius with $R\le R_1$, 
while the agent moves with reduced directional persistence with $\Drot=D_2>0$ for $R>R_1$.
This strategy exploits a generic scattering effect that provides ABPs with multiple search attempts \cite{Kromer2020}. 
Like in strategy (i-a), inner search is terminated after an inner search time $\tau$, after which the agent resumes outer search 
with only ballistic motion until proximity of the next target is detected .
Below we show that the rate of target encounter is increased for adaptive inner search
(black curves in Fig.~\ref{figure2}).
\end{enumerate}

\paragraph{(i-a) Fixed inner search time.}
We introduce a strict \textit{timer} for inner search:
The ABP automatically resets its internal state back to outer search after a maximal inner search time $\tau$ 
(except in the unlikely event that the agent finds itself again in a target zone defined by $R\le R_1$). 
This defines a \textit{composite search} strategy \cite{Nolting2015}.

What is the optimal time $\tau_\mathrm{opt}$ the ABP should spend in `inner search mode' 
before switching back to ballistic motion (`giving-up time' \cite{Nolting2015}),
in order to maximize the rate $k$ of target encounters?
The answer depends on the density of targets $\rho$ 
and the probability $p_\mathrm{in}(\tau)$ of successful inner search within time $\tau$. 
The derivative
$\mathcal{F}(\tau)=d p_\mathrm{in}(\tau)/d\tau$ 
denotes the (non-normalized) distribution of first-passage times for inner search.

For composite search consisting of inner and outer search, 
we have 
$k = k_\mathrm{out}\,p_\mathrm{in}(\tau)$, 
where $k_\mathrm{out}$ denotes the rate at which the ABP enters the target zone around some target.
For ballistic motion, $k_\mathrm{out}$ equals the product of $\rho$ 
and the effective search volume per unit time
$\sigma = \pi R_1^2\, v$ \cite{Rothschild1949}.
Thus,
$k_\mathrm{out}\approx \rho\, \sigma (1-k_\mathrm{out}\tau)$,
where we take into account the time fraction $k_\mathrm{out}\tau$ spent in inner search.
We neglect the small probability that the ABP is again inside a target zone at the end of inner search.
This expression 
can be rewritten as
$k_\mathrm{out} = [ (\rho\,\sigma)^{-1} + \tau ]^{-1}$.
The optimal duration $\tau_\mathrm{opt}$ of inner search that maximizes $k(\tau)$
must satisfy
$0=\partial k/\partial \tau_{|\tau=\tau_\mathrm{opt}}$, 
which we can re-arrange as
\begin{equation}
\label{eq:topt}
\frac{\partial}{\partial \tau} \ln p_\mathrm{in}(\tau)_{|\tau=\tauopt}
=
[(\rho \, \pi R_1^2 \, v)^{-1}+\tau_\mathrm{opt}]^{-1}\quad.
\end{equation}

Eq.~(\ref{eq:topt}) embodies the trade-off 
between continuation of inner search and decampment to new targets
in the spirit of a \textit{Marginal Value Theorem} \cite{Charnov1976}. 
If targets are denser, less time should be spent in inner search, 
see Fig.~\ref{figure2}(b), red curves.
For long times $\tau \Drot\gg 1$, persistent random walks can be approximated by diffusion.
This implies an asymptotic Sparre-Anderson scaling
$dp_\mathrm{in}/d\tau\sim \tau^{-3/2}$ 
\cite{FellerII}.
In the limit of sparse targets, 
where the volume fraction occupied by the target zones of all targets is much smaller than one,
$\rho V_1\ll 1$, 
a straight-forward but lengthy calculation gives
(see Supplemental Material (SM) text) 
\begin{equation}
\label{eq:tau_opt}
\tau_\mathrm{opt}\sim\rho^{-2/3}
\quad.
\end{equation}
The ABP will spend considerably more time in outer search compared to inner search, $k_\mathrm{out}\tau_\mathrm{opt}\ll 1$.
Together with 
$p_\mathrm{in}(\tau_\mathrm{opt}) 
\lesssim \lim_{\tau\rightarrow\infty} p_\mathrm{in}(\tau)
\approx R_0/R_1$, 
we conclude for the rate of target encounter for composite search, 
$k \approx \rho\,\pi R_0 R_1\, v$.
We may re-state this result equivalently as follows: 
composite search with fixed inner search time  according to strategy (i-a) 
increases the effective cross-sectional area of target encounter to
$k/(\rho v) \lesssim \pi R_0 R_1$, 
compared to 
$k/(\rho v) = \pi R_0^2$
for pure ballistic motion.
A detailed mathematical derivation can be found in SM text.

\paragraph{(i-b) Stochastic timer.}
Above we considered the case of a fixed inner search time $\tau$ for sake of simplicity, 
yet real artificial microswimmers will likely not be able to control inner search time precisely. 
Instead, we expect that the microswimmer can only realize a distribution $q(\tau)$ of inner search times that vary around a mean value $\ol{\tau}$.
In the possibly simplest example, switching back to outer search mode could be controlled by a 
single-rate process with rate constant $1/\ol{\tau}$;
in this case, inner search times $\tau$ follow an exponential distribution with mean $\ol{\tau}$
\begin{equation}
\label{eq:qbar}
q_\ol{\tau}(\tau) = \ol{\tau}^{-1} \exp(-\tau/\ol{\tau})\quad.
\end{equation} 
All results derived above hold also in this case
(and in fact even for $q_\ol{\tau}(\tau)=\ol{\tau}^{-1} f(\tau/\ol{\tau})$ with some arbitrary function $f$), 
provided the probability $p_\mathrm{in}(\tau)$ of successful inner search is replaced by 
an effective probability distribution 
$\ol{p}_\mathrm{in}(\ol{\tau}) = \int_0^\infty \!d\tau\, p_\mathrm{in}(\tau) q_\ol{\tau}(\tau)$.
In particular, we have the asymptotic scaling
$d\ol{p}_\mathrm{in}/d\ol{\tau}\sim \ol{\tau}^{-3/2}$, and
Eq.~(\ref{eq:tau_opt}) holds analogously for the optimal mean inner search time $\ol{\tau}$, 
see Fig.~\ref{figure2}, blue curves. 

\paragraph{(ii) Adaptive inner search.}
The probability for successful inner search $p_{\text{in}}(\tau)$ can be further increased by choosing an even more efficient inner search strategy
\cite{Kromer2020}. 
Instead of using a constant rotational diffusion coefficient $\Drot$ during inner search,
the agent may employ a position-dependent rotational diffusion coefficient $\Drot(\x)$ during inner search and 
switch its rotational diffusion coefficient between two values $D_1$ and $D_2$ inside and outside a target zone of radius $R_1$ around the target, respectively. 
If $D_1\ll D_2$, 
this adaptive inner search strategy increases the probability to eventually find the target substantially 
by exploiting a dynamic scattering effect \cite{Kromer2020}. 
In short, ABP that head away from the target will switch to motion with low directional persistence as soon as they move beyond distance $R_1$,
which results in a high probability to reverse direction and enter the target zone again. 
There, the ABP resumes ballistic motion, providing the ABP with another attempt to hit the target. 
In the following, we set $D_1=0$ for the rotational diffusion coefficient $\Drot$ inside the target zone.

\begin{figure}
\includegraphics[width=\linewidth]{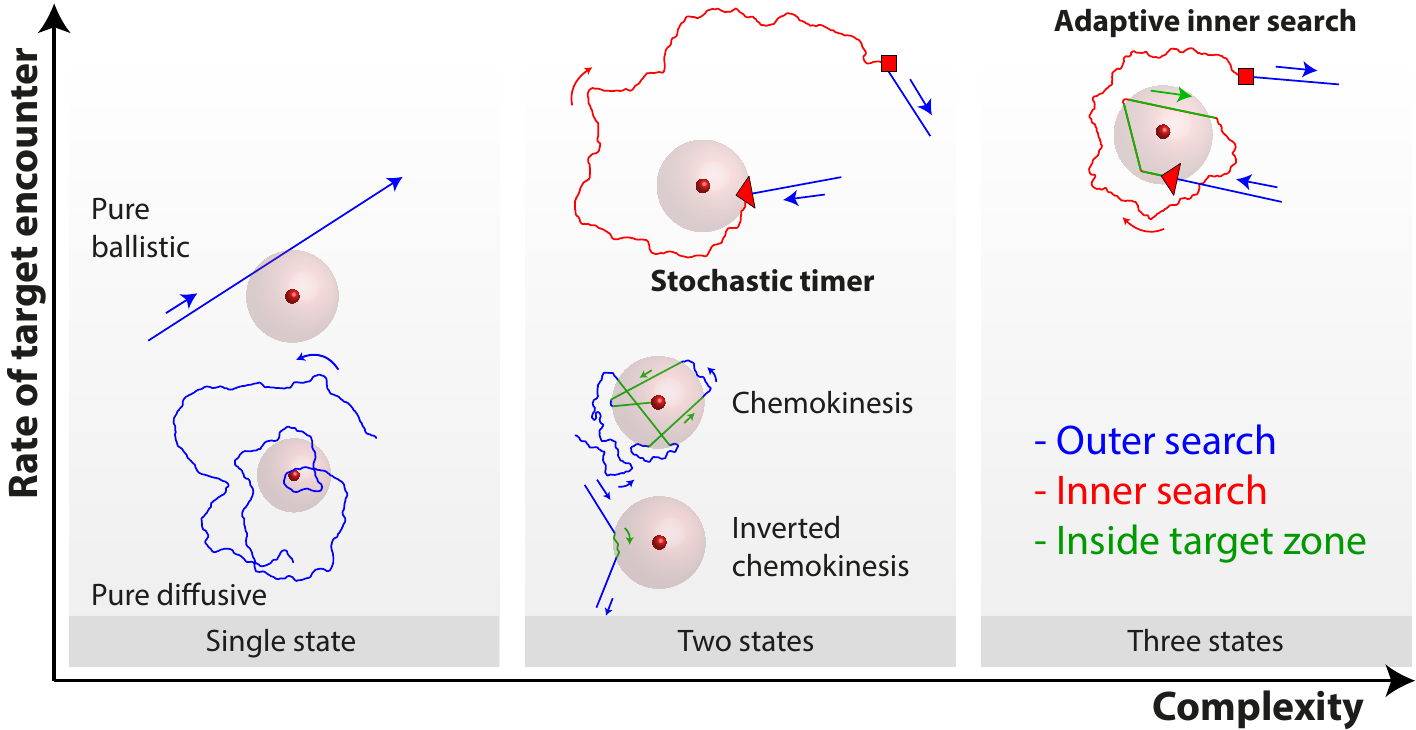}
\caption{
\textbf{Comparison of random search strategies.} 
Shown are schematics of the different search strategies.
In addition to the two strategies studied in this manuscript, 
composite search with \textit{timer} for inner search, and
\textit{adaptive} inner search studied in this manuscript, 
\textit{pure ballistic motion} (with $\Drot=0$) and 
\textit{pure diffusive motion} (with $\Drot\gg v/R_0$), as well as
instantaneous \textit{chemokinesis}, and
\textit{inverted chemokinesis} 
are shown.
For instantaneous chemokinesis, the ABP changes its rotational diffusion coefficient instantaneously as function of target distance, 
with $\Drot=D_1$ inside the target zone (ros{\'e}, target distance $R\le R_1$, green trajectory), and
$\Drot=D_2$ outside the target zone ($R>R_1$, blue trajectory).
The case $D_1=0$ and $D_2\gg v/R_0$ labeled \textit{chemokinesis} exploits a beneficial inward scattering effect, 
whereas the case $D_1\gg v/R_0$ and $D_2=0$ labeled \textit{inverted chemokinesis} suffers from disadvantageous outward scattering \cite{Kromer2020}.
The different strategies are positioned according to a qualitative assessment of 
their control complexity and search performance.
SM text contains quantitative results for the limit of low target density and high rotational diffusion coefficient.
}
\label{figure3}
\end{figure}

An analytical calculation shows that the normalized rate of target encounter, $k/(v\rho)$, 
which represents an effective cross-sectional area of targets, 
equals $\pi R_1^2$ for adaptive search 
in the limit of sparse targets ($\rho V_1\ll 1$) and large rotational diffusion coefficient ($D_2\gg v/R_0$), 
see SM text for details. 
For comparison, 
we had found lower effective cross-sectional areas 
$k/(v\rho)\lesssim \pi R_0 R_1$ for inner search with constant $\Drot\gg v/R_0$, and 
$k/(v\rho) = \pi R_0^2$ for pure ballistic motion above, see also Fig.~\ref{figure3}.

\paragraph{Numerical results confirm scaling law.}

Fig.~\ref{figure2}(a) shows simulation results for the first passage time density $\mathcal{F(\tau)}$ to find the target exactly after time $\tau$ 
if the ABP starts at distance $R_1$ 
(with random initial direction as in \cite{Kromer2020}). 
For moderate values of the rotational diffusion coefficient $\Drot$,
all three inner search strategies introduced above, i.e.,
inner search with \textit{fixed timer} (red curves), 
inner search with \textit{stochastic timer} (blue curves), and 
adaptive inner search (black curves), 
give rise to similar distributions $\mathcal{F(\tau)}$.
For a high rotational diffusion coefficient, however,
the distribution $\mathcal{F(\tau)}$ is substantially higher for adaptive inner search.
In all cases, numerical results confirm the asymptotic scaling $\mathcal{F}(\tau)\sim \tau^{-3/2}$.

Fig.~\ref{figure2}(b) displays the corresponding optimal inner search time $\tauopt$,
which was numerically determined from Eq.~(\ref{eq:topt}).
Generally, strategies for which the density of first passage times is comparably higher for short inner search times $\tau$
display shorter optimal inner search times $\tauopt$. 
Intuitively, several short inner search episodes are more efficient than fewer longer ones.
All cases confirm the asymptotic scaling $\tauopt\sim \rho^{-2/3}$.

\paragraph{Discussion.}

We studied optimal search strategies for hidden targets,
whose presence can be sensed at a distance.
Using a minimal model of a chemokinetic agent 
that can switch between different rotational diffusion coefficients, 
we explain how search agents employing composite search
find targets more efficiently if they posses memory
in the form of an internal state
(`outer search mode' versus `inner search mode'). 

Intuitively, the improved search success can be explained as follows:
if targets are randomly distributed in an infinite search domain (and can only be visited once), 
the optimal search strategy that maximizes the rate of target encounter 
is ballistic motion, i.e., moving simply along a straight path \cite{Viswanathan1999, Bartumeus2002, Tejedor2012}. 
If a search agent, however, receives the information that a target must be close, 
it is advantageous to transiently switch to less persistent motion to search the local neighborhood more thoroughly.
Specifically, if a target of radius $R_0$ is known to be at most a distance $R_1$ away, 
the probability to eventually find this target using Brownian motion is $p^\ast(R_0|R_1)=R_0/R_1$ if search time is unlimited
\cite{Berg:random}.
Mathematically, the conditional mean-first passage time to find the target diverges in this case, 
but this is not a practical limitation for search agents that use a finite inner search time. 
Indeed, a small fraction of eventually succesful realizations will require a long time to reach the target 
(power-law tail of the distribution of conditional first-passage times). 
Yet, most of those realizations that reach the target will have done so already within a time-scale $\sim R_1^2/D_\mathrm{eff}$, 
where $D_\mathrm{eff}=v_0^2/(6\Drot)$ is the effective translational diffusion coefficient of the search agent. 
Thus, the search agent faces a trade-off choice: 
shall it continue the inner search for a particular target that is known to have been close, 
to get closer to the best achievable probability $\lim_{\tau\rightarrow\infty} p_{\text{in}}(\tau)$ to find this particular target?
Or rather give up inner search, wander off for good, 
and try to find other targets, to ultimately maximize its rate $k$ of target encounters?
This trade-off choice defines an optimal time $\tauopt$ for inner search as function of target density:
if the density $\rho$ of targets is higher, it pays off to abandon inner search earlier,
$\tauopt\sim\rho^{-2/3}$.

Our generic framework allows to investigate different combinations of outer and inner search strategies, 
see also Fig.~\ref{figure3} and the SM.
In general, strategies with ballistic motion during outer search outperform strategies with effective diffusive motion.
For ballistic outer search, 
different inner search strategies differ in search performance 
(quantified in terms of an effective cross-sectional area of targets in the SM), 
with adaptive inner search performing best. 
Future work will address the effect of measurement noise in target zone detection.
More complex strategies can be realized 
if agents respond in a gradual manner to the distance from the nearest target, 
e.g., in the form of distance-dependent swimming speed \cite{Vuijk2018}, 
distance-dependent rotational diffusion coefficient \cite{Kromer2020}, 
or for dimers of elastically coupled active and passive particles in a spatial activity field \cite{Vuijk2020}. 

Search strategies, 
such as the adaptive inner search strategy presented here,
where agents increase their directional persistence when they move towards the target, 
are conceptually related to the principle of run-and-tumble chemotaxis, 
where changes in a temporal concentration signal regulate the strength of directional fluctuations \cite{Berg1972}. 
Yet our minimal implementation is much simpler:
in contrast to chemotaxis, it suffices if the agent compares local concentration of signaling molecules diffusing from a target to a fixed threshold.

Chemotactic navigation could additionally increase 
the probability to find the target when the search agent is already very close to the target and chemical gradients can be measured with sufficient precision \cite{berg1977, bialek2005physical, mora2019physical, Novak2021,Lange2019}.
We speculate that composite random search as described here
might have been an evolutionary precursor of chemotaxis strategies that additionally require sensory adaption. 
Composite random search strategies without sensory adaptation may represent a promising route for the design of future active microswimmers 
that need to find hidden targets.


\begin{acknowledgments}
AA acknowledges support by the DFG (FR3429/3-1);
BMF acknnowleges support through the Heisenberg program of the DFG (FR3429/4-1);
JAK, AA, and BMF were supported through the Excellence Initiative by the German Federal and State Governments 
(Clusters of Excellence cfaed EXC-1056 and PoL EXC-2068).

We would like to dedicate this manuscript to the memory of Lutz Schimansky-Geier.
\end{acknowledgments}



\cleardoublepage

\appendix

\section{Supplemental Material}

{\noindent
Justus A. Kromer, Andrea Auconi,
Benjamin M. Friedrich: 
\textbf{
Composite search of active particles in three-dimensional space based on non-directional cues
}}

\renewcommand{\theequation}{S\arabic{equation}}    
\setcounter{equation}{0}  
\renewcommand{\thefigure}{S\arabic{figure}}    
\setcounter{figure}{0}  
\renewcommand{\thetable}{S\arabic{table}}    
\setcounter{table}{0}  

\subsection{Numerical methods}

For numeric integration of Eqs. (1), (2) for the inner search problem, 
we used an explicit Euler-Maruyama method with integration time step $\Delta t=2 \times 10^{-3}\,R_0/v$.
ABPs were initially positioned at $R_1$ with initial direction angle $\psi$ distributed according to 
$p(\psi,t=0)=\sin(2 \psi)$ for $\psi \in [0, \pi/2]$ \cite{Kromer2020}. 
Simulations were stopped after a maximum search time of $2 \times 10^4\,R_0/v$ 
(corresponding to a maximum trajectory length of $10^3\,R_1$). 
Consistent results were obtained in preliminary simulations of ABP trajectories in three-dimensional space 
using an Euler-Heun scheme with matrix exponentials for propagation of the Frenet-Serret frame.

To compute the effective probability distribution 
$\ol{p}_\mathrm{in}(\ol{\tau}) = \int_0^\infty \!d\tau\, p_\mathrm{in}(\tau) q_\ol{\tau}(\tau)$
for search agents with stochastic timer [blue curves in Fig.~\ref{figure2}], 
we used simulation results for 
$p_\mathrm{in}(\tau)$ for $0\le\tau\le t_\mathrm{max}$ with $t_\mathrm{max}=5 \times 10^4\,\tau_0$, 
and extrapolated $\mathcal{F}(\tau) = c\, \tau^{-3/2}$ for $\tau>t_\mathrm{max}$ 
with prefactor $c$ determined by a fit. 

\subsection{Analytic derivation of scaling law}

We can mathematically prove the observed power law scaling of the optimal inner search time $\tau_\mathrm{opt}$
in the limit of low directional persistence, $\Drot \gg v/R_0$.
In this limit, 
we can approximate the motion of the agent for inner search by diffusive motion with effective translational diffusion coefficient 
$\Deff=l_p v/3=(2/3)\,v R_0$ 
(corresponding to the example of an effective translational diffusion coefficient of an ABP 
with rotational diffusion coefficient $\Drot = v/(4 R_0)$ and persistence length $l_p=v/(2\Drot)=2 R_0$
as used in Fig.~\ref{figure2}). 
For diffusive motion,
the probability to find a spherical target of radius $R_0$ within time $\tau$ if starting at $R=R_1$, 
is given by
\begin{equation}
\label{eq:pin_diff}
p_\mathrm{in}(\tau) = \frac{R_0}{R_1}\, \left[ 1- \mathrm{Erf} \left( \frac{R_1-R_0}{\sqrt{4\Deff\tau}}\right) \right]
\quad.
\end{equation} 

We include the derivation of Eq.~(\ref{eq:pin_diff}) for completeness:
we consider the diffusion equation in spherical coordinates 
for a radially symmetric probability density $p(R,t)$ for position in three-dimensional space (with units of an inverse volume)
\begin{equation}
\label{radial_diff_eqn}
\frac{\partial}{\partial t}{p}=\Deff\,R^{-1}\,\frac{\partial^2}{\partial R^2}\,\left( R \, p \right)
\quad, 
\end{equation}
with initial condition
$p(R,0)=(4\pi R_1^2)^{-1}\, \delta(R-R_1)$,
and absorbing boundary conditions at $R=R_0$,
$p(R_0,t)=0$,
and use the method of images to find the time-dependent solution
\begin{align*}
p(R,t)= \frac{1}{4\pi R_1^2\,\sqrt{4\pi \Deff t}} \frac{R_1}{R} 
\Big[ 
 & \exp\left(-\frac{(R-R_1)^2}{4\Deff t}\right) \\
-& \exp\left(-\frac{(R+R_1-2 R_0)^2}{4\Deff t}\right)
\Big] \quad.
\end{align*}
This solution satisfies the absorbing boundary condition $p(R_0,t)=0$ at $R=R_0$.
Now, 
$p_\mathrm{in}(\tau) = \int_0^\tau dt\, \mathcal{F}(t)$, 
where the probability current $J(t)$ at time $t$ 
to the absorbing sphere of radius $R_0$ at the origin is given by
$\mathcal{F}(t) = 4\pi\,R_0^2\,\Deff\,\partial p(R,t)/\partial R_{|R=R_0}$.

Eq.~(\ref{eq:pin_diff}) becomes exact in the limit $l_p\ll R_0$.
Note that a similar result, yet with additional prefactor $R_0 v/(4\Deff)$, 
was derived for a case of intermediate persistence length $l_p$ 
with $R_0\ll l_p\ll R_1$ (or $R_0 v\ll \Deff\ll R_1 v$) \cite{Friedrich2008}
\begin{equation}
\label{eq:pin_diff2}
p_\mathrm{in}(\tau) = \frac{R_0 v}{4\Deff}\,\frac{R_0}{R_1}\, \left[ 1- \mathrm{Erf} \left( \frac{R_1}{\sqrt{4 \Deff \tau}}\right)\right]
\quad.
\end{equation} 
Here, $\Deff=l_p v/3$ denotes the effective translational diffusion coefficient of a persistent random walk in three-dimensional space with general persistence length $l_p$.

For the rate $k_\mathrm{out}$ at which the ABP enters the target zone around some target, 
we have
\begin{align}
\label{eq:k_out}
k_\mathrm{out} = \rho v\,\pi R_1^2 (1-k_\mathrm{out}\tau) 
= \frac{\rho v\,\pi R_1^2}{1 + \rho v\,\pi R_1^2\,\tau} \quad.
\end{align}
By combining Eq.~(\ref{eq:pin_diff}) and Eq.~(\ref{eq:k_out}), 
we obtain for the rate $k$ of target encounter
\begin{align}
k 
& = k_\mathrm{out}\,p_\mathrm{in}
= \frac{\rho v\,\pi R_0 R_1}{1 + \rho v\,\pi R_1^2\,\tau} \,
\left[ 1- \mathrm{Erf} \left( \frac{R_1-R_0}{\sqrt{4\Deff\tau}}\right) \right]
\quad.
\end{align}
In the limit of small targets $R_0\ll R_1$, 
we can rewrite this expression as 
\begin{align}
k &= \rho\,\pi\,R_0 R_1\, v\, \frac{\beta^2}{\beta^2+\Lambda} \left[ 1-\mathrm{Erf}(\beta) \right]
\quad,
\end{align}
where we introduced short-hand
\begin{equation}
\Lambda = \frac{3\pi}{8}\,\frac{R_1}{R_0}\,\rho R_1^3, \quad
\beta^2 = \frac{3}{8}\,\frac{R_1}{v\tau}\,\frac{R_1}{R_0}=\frac{R_1^2}{4\Deff \tau}
\quad.
\end{equation}

In the limit $\Lambda\ll 1$, corresponding to $\rho V_1\ll R_0/R_1$, 
we find that $k$ becomes maximal for 
$\beta_\mathrm{opt} = \pi^{1/6}\,\Lambda^{1/3}$.
Thus, the optimal inner search time $\tau_\mathrm{opt}$ reads
\begin{equation}
\tau_\mathrm{opt} = \frac{3^{1/3}}{2\pi}\,\frac{R_0}{v}\,(\rho\,R_0^2 R_1)^{-2/3}\quad.
\end{equation} 
Hence, 
$\tau_\mathrm{opt}$ 
decreases if $R_0$ or $R_1$ become larger, and
increases if targets become sparser.

In the limit $\rho V_1\ll R_0/R_1$ with $\beta\ll 1$, 
the success probability of inner search approaches the limit value for time-unrestricted diffusive search,
$\lim_{\tau\rightarrow\infty} p_\mathrm{in} = R_0/R_1$.
The time fraction spent in inner search mode is negligible in this limit, 
$k\tau \approx (R_0^2 R_1 \rho)^{1/3}\ll 1$. 
For the rate $k$ of target encounter, we find 
\begin{equation}
k \approx \rho\,\pi\, R_0 R_1 \, v
\quad.
\label{eq:k_fixed}
\end{equation}
This result corresponds to an effective cross-sectional area of the target 
$k/(v\rho) \approx \pi R_0 R_1$
for the case of composite search with fixed inner search considered here (in the appropriate limit case), 
see also Table~\ref{table:k}.

\newcommand{\kdiff}{k_\mathrm{diff}}

\subsection{Reference case: Global diffusive search}
As a reference case, we additionally consider a case of effective diffusive search without time-restriction, and
a high rotational diffusion coefficient $\Drot \gg v/R_0$ that is constant in time and space.
As always, we consider search in three-dimensional space for randomly distributed non-revisitable targets.
A calculation analogous to the derivation of Eq.~(\ref{eq:k_fixed}) yields the steady-state target encounter rate $\kdiff$ 
for this case (time-unrestricted diffusive search for non-revisitable targets in three-dimensional space).
This rate is well defined only in dimension $d\ge 3$, 
which follows from P\'olya's classical result on time-discrete random walks on a $d$-dimensional lattices \cite{Dvoretzky1951}.

We first consider a purely diffusive search agent with effective translational diffusion coefficient $\Deff$ in three-dimensional space, as well as 
Poisson distributed spherical targets with radii $R_0$.
We switch to a coordinate frame that is co-moving with the diffusive agent.
In this frame, the centers of the targets are diffusing and become absorbed 
once they reach a spherical shell of radius $R_0$ concentric with the agent.
We make the simplifying assumption that the diffusive motion of the target centers can be considered independent.
The probability density $p_\mathrm{target}(R,t)$ of those target centers that have not yet been absorbed 
is radially symmetric (with units of an inverse volume) 
and obeys the diffusion equation Eq.~(\ref{radial_diff_eqn}). 
The steady-state probability density reads 
\begin{equation}
p_\mathrm{target}^\ast (R) = \rho \, \left( 1 - \frac{R_0}{R} \right)\quad,
\end{equation}
satisfying the boundary conditions
$p_\mathrm{target}^\ast (R_0)=0$ and
$\lim_{R\rightarrow\infty} p_\mathrm{target}^\ast (R)=\rho$.
Thus, we find for the steady-state current of target centers to the absorbing shell
\begin{align}
\kdiff &=4\pi R_0^2\, \Deff\, \frac{\partial}{\partial R} p_\mathrm{target}^\ast (R) _{|R=R_0} \\
&= \rho\, 4\pi R_0\, \Deff\quad. 
\end{align}

An ABP with high rotational diffusion coefficient $\Drot \gg v/R_0$ can be approximately described as a diffusive particle
with effective translational diffusion coefficient $\Deff = l_p v/3$,
where $l_p=v/(2\Drot)$ denotes the persistence length of the ABP. 
Thus, 
\begin{equation}
\frac{\kdiff}{\rho v} \approx \frac{4\pi}{3}\, R_0 l_p\quad.
\label{eq:k_diff}
\end{equation}

\subsection{Limit values of target encounter rate}

We report limit values of the target encounter rate $k$ for different 
inner search strategies in the limit of large rotational diffusion coefficients, see table \ref{table:k}.
Target encounter rates are reported as an effective cross-sectional area of the targets, $k/(v\rho)$.
For the case of fixed inner search time with effective diffusive motion in inner search (first column), 
the target encouter rate equals $k=k_\mathrm{out} p_\mathrm{in}$, 
where 
$k_\mathrm{out}/(v\rho) = \pi R_1^2$ 
is the normalized rate of target encounter for ballistic outer search, and 
$p_\mathrm{in} \approx R_0 / R_1$
the success probability of inner search. 
For the case of adaptive inner search 
[with ballistic motion for $R\leq R_1$ and effective diffusive motion for $R>R_1$, strategy (ii)]
(second column), inward scattering of outgoing trajectories implies $p_\mathrm{in} \lesssim 1$ \cite{Kromer2020}. 
For comparison, we also report the case of purely ballistic search, 
where the ABP employs $\Drot = 0$ both for outer and inner search (third column);
in this case, 
$k/(v\rho) = \pi R_0^2$. 
Finally, we consider a case of inverted adaptive search, 
where the ABP employs a high rotational diffusion coefficient $\Drot=D_1\gg v/R_0$ inside the target zone defined by $R\le R_1$, 
but ballistic motion outside, $\Drot = D_2 = 0$ (fourth column). 
In this case, 
$p_\mathrm{in} \approx 0$ 
as consequence of an outward scattering effect of ingoing trajectories \cite{Kromer2020}.

\begin{table*}[b]
\begin{center}
\begin{tabular}{|c|c|c|c|}
\hline
\multicolumn{4}{|c|}{Strategy for \textit{inner search} (\textit{outer search}: ballistic)}\\[1mm]
\hline
Fixed inner search time\ &\ Adaptive inner search\ &\ Ballistic inner search\ &\ Inverted adaptive inner search\ \\[1mm]
\hline
$\pi R_0 R_1$ & $\pi R_1^2$ & $\pi R_0^2$ & $0$ \\[1mm]
\hline
\end{tabular}
\end{center}
\caption[]{
Normalized rate of target encounter for different inner search strategies
reported as $k/(v\rho)$ 
with units of an effective cross-sectional area of targets.
We assume ballistic motion with $\Drot=0$ in outer search throughout. 
\textbf{First column:}
we consider diffusive motion with constant rotational diffusion coefficient $\Drot \gg v/R_0$, 
corresponding to the strategy (i-a) of \textit{fixed inner search time} considered in the main text. 
Analogous results are found for the strategy (i-b) with \textit{stochastic timer} with random inner search times. 
\textbf{Second column:}
we consider \textit{adaptive inner search}, strategy (ii),
with position-dependent rotational diffusion coefficient $\Drot(\x)$ ($D_1=0$ and $D_2\gg v/R_0$) as considered in the main text.
\textbf{Third column:}
additionally, we state the result for an ABP that moves always ballistically, 
i.e., for which outer and inner search are indistinguishable. 
\textbf{Fourth column:}
finally, we consider an inverted adaptive inner search strategy, 
where the position-dependent rotational diffusion coefficient $\Drot(\x)$ 
equals $D_1\gg v/R_0$ for $R\leq R_1$ and $D_2=0$ for $R>R_1$.
}
\label{table:k}
\end{table*}

Lastly, one may ask about composite search strategies that employ effective diffusive motion in outer search, 
with high rotational diffusion coefficient $\Drot$. 
In this case, a composite search with constant rotational diffusion coefficient during inner search 
(e.g., \textit{fixed inner search time}) 
does not provide any considerable benefit.
Only adaptive inner search considerably increases the rate of target encouter, yielding 
$k/(v\rho) \approx (4\pi/3)\,R_1 l_p$,
where 
$D_2=v/(2l_p)\gg v/R_0$
is the rotational diffusion coefficient used during inner search outside the target zone.
(Here, we used Eq.~(\ref{eq:k_diff}) for $k_\mathrm{out}$, replacing $R_0$ by $R_1$, and $p_\mathrm{in}\approx 1$.)
However, this rate is still smaller than the corresponding rate for adaptive inner search and ballistic outer search, 
see~Table~\ref{table:k}.

\end{document}